\documentstyle[preprint,aps,prl]{revtex}
\begin{document}
\preprint{ NSFITP-93-100}
\draft
\title{
Weak-coupling expansions for the attractive Holstein and Hubbard models
}
\author{
J. K. Freericks\cite{Davis} and D. J. Scalapino.
}
\address{
Institute for Theoretical Physics,
and
Department of Physics,\\
University of California,
Santa Barbara, CA 93106
}
\date{\today}
\maketitle
\widetext
\begin{abstract}
Weak-coupling expansions (conserving approximations) are carried out
for the attractive
Holstein and Hubbard models (on an infinite-dimensional hypercubic
lattice) that include all bandstructure and vertex
correction effects.
Quantum fluctuations are found to renormalize transition temperatures
by factors of order unity, but may be incorporated
into the superconducting channel of
Migdal-Eliashberg theory by renormalizing the phonon
frequency and the interaction strength.
\end{abstract}
\narrowtext
Interacting electron-phonon systems sustain both charge-density wave (CDW)
and superconducting (SC) order.  Migdal \cite{migdal} and Eliashberg
\cite{eliashberg} showed that a self-consistent theory for the
electron-phonon problem can be constructed that neglects so-called
vertex corrections (Migdal's theorem) in the limit were the phonon frequency
is small compared to the fermi energy.
Numerical calculations of SC transition temperatures based upon
experimentally extracted (interacting) electron-phonon spectral
densities \cite{mcmillan}
soon followed.  The Migdal-Eliashberg (ME) theory is successful for
predicting transition temperatures (and other materials properties) of
virtually all low temperature superconductors \cite{reviews,wu}.

However, ME theory assumes that the electronic bandwidth is infinite, and that
vertex corrections may be completely ignored.  Although the former can be
handled by simple modifications of the basic theory \cite{reviews,marsiglio},
the latter require complicated extensions of the ME framework.
Grabowski and Sham \cite{sham} carried out such an analysis for the
interacting electron
gas and found that vertex corrections strongly suppressed $T_c$ for
plasma frequencies larger than ten percent of the fermi energy.  Van Dongen
\cite{vandongen} analyzed the Hubbard model (which can be viewed as the
infinite-frequency limit of an electron-phonon model) and showed that
quantum fluctuations (vertex corrections)
reduced Hartree-Fock transition temperatures
by more than a factor of three.

In this contribution, ME theory is extended to include vertex corrections and
produce weak-coupling expansions
for $T_c$ that are accurate for the entire
range of phonon frequencies (Coulomb repulsion effects are neglected
here).  Comparison of these weak-coupling expansions
are made to quantum Monte Carlo simulations over a wide range of parameter
space.  Effects due to vertex corrections are strong when expressed in terms
of the bare microscopic parameters, but can be incorporated into the ME theory
(to lowest order) by renormalizing these parameters.

The electron-phonon Hamiltonian chosen here is that of the Holstein model
\cite{holstein} in which conduction electrons interact with localized
(Einstein) phonons:
\begin{equation}
  H = - {{t^*}\over{2\sqrt{d}}} \sum_{\langle j,k\rangle \sigma} (
   c_{j\sigma}^{\dag }
  c_{k\sigma} + c_{k\sigma}^{\dag  } c_{j\sigma} ) + g\sum_j  x_j
  (n_{j\uparrow}+n_{j\downarrow}) + {{1}\over{2}} M \Omega^2 \sum_j x_{j}^2 +
  {{1}\over{2M}} \sum_j p_{j}^2
\label{eq: holham}
\end{equation}
where $c_{j\sigma}^{\dag}$ ($c_{j\sigma}$) creates (destroys) an electron at
site $j$ with spin $\sigma$, $n_{j\sigma}=c_{j\sigma}^{\dag}c_{j\sigma}$ is
the electron number operator, and $x_j$ ($p_j$) is the phonon coordinate
(momentum) at site $j$.  The hopping matrix elements connect the nearest
neighbors of a hypercubic lattice in $d$-dimensions and for concreteness
the infinite-dimensional limit $(d\rightarrow\infty)$ of Metzner and
Vollhardt \cite{metzandvol}
is taken ($1/d$ corrections should be less than 5\% in three dimensions
\cite{vandongen}).  The unit of energy is chosen to be the rescaled matrix
element $t^*$.  The phonon has a mass $M$ (chosen to
be $M=1$), a
frequency $\Omega$, and a spring constant $\kappa\equiv M\Omega^2$ associated
with it.  The electron-phonon coupling constant is denoted by $g$; the
effective electron-electron interaction strength is then the bipolaron
binding energy
\begin{equation}
  U\equiv - {{g^2}\over{M\Omega^2}}=-{g^2\over\kappa} \quad .
 \label{eq: udef}
\end{equation}
In the limit where $\kappa$ remains finite and $\Omega$ is large compared to
the bandwidth $(\Omega\rightarrow\infty ,U={\rm
finite})$, the Holstein model maps onto the attractive Hubbard model
\cite{hubbard}
\begin{equation}
  H = - {{t^*}\over{2\sqrt{d}}} \sum_{\langle j,k\rangle \sigma} (
   c_{j\sigma}^{\dag }
  c_{k\sigma} + c_{k\sigma}^{\dag  } c_{j\sigma} ) + U \sum_j
  n_{j\uparrow}n_{j\downarrow}
\label{eq: hubbham}
\end{equation}
with $U$ defined by Eq.~(\ref{eq: udef}).

Our weak-coupling expansions are based upon the conserving approximations of
Baym and Kadanoff \cite{baym}: the free energy $\Phi$ is expanded to
a given order in the interaction strength; the self energy
$\Sigma(i\omega_n)$ is determined by functional differentiation
$\Sigma(i\omega_n)=\delta\Phi/\delta G(i\omega_n)$ at each Matsubara frequency
$\omega_n\equiv (2n+1)\pi T$; and the irreducible vertex functions
$\Gamma(i\omega_m,i\omega_n)$ (in the relevant channels) are determined by
a second functional differentiation.  The free energy must be expanded to order
$U^2$ in order to determine the correct transition temperature in the limit
$|U|\rightarrow 0$\cite{vandongen}.  The necessary diagrams for the free
energy in the Holstein model are presented in Fig.\ \ref{fig:1}.  Note
that conventional
ME theory is identical to a first-order (Hartree-Fock) conserving approximation
for the SC channel (with the exception that the phonon propagator is dressed),
but is quite different in the CDW channel, because exchange diagrams are
important even at the Hartree-Fock level \cite{baym}.

The many-body problem in infinite dimensions maps onto a self-consistently
embedded Anderson impurity problem because the self-energy has no momentum
dependence \cite{georgesandkotliar,jarrell}.  The
self-consistent embedding of the impurity problem is solved
by iteration \cite{jarrell}:
the self energy is calculated from the dressed (local) Green's functions
by a conserving approximation for the impurity problem $\Sigma [G]$
and the new (local) Green's function is determined by the embedding
\begin{equation}
  G(i\omega_n)=\int_{-\infty}^{\infty}dy{\rho(y)\over
i\omega_n+\mu-\Sigma_n[G]-y}
\equiv F_{\infty}(i\omega_n+\mu-\Sigma_n[G])
\label{eq: embed}
\end{equation}
with $\rho (y)\equiv \exp(-y^2)/\sqrt{\pi}$ the bare density of states in
infinite dimensions.

To illustrate the connection between the microscopic parameters of the
Holstein model and the renormalized parameters of the ME theory, the SC
transition temperature is calculated in a square-well approximation at
half-filling.
The transition temperature is assumed to satisfy $T_c << \min (\Omega,t^*)$.
In
this case, the electron self energy becomes (to lowest order in $U$)
\cite{baym}
\begin{equation}
\Sigma(i\omega_n)\equiv i\omega_nZ(i\omega_n)\quad ; \quad Z(i\omega_n)=1-U
\int_0^{\infty}dy \rho(y){\Omega\over \omega_n^2+(\Omega+y)^2}
\label{eq: zdef}
\end{equation}
and the irreducible vertex function in the SC channel satisfies
(to second-order in $U$) \cite{baym}
\begin{eqnarray}
\Gamma_{SC}(i\omega_m,i\omega_n)&=&UT\{\theta(\omega_c-|\omega_m|)
\theta(\omega_c-|\omega_n|)[1+{2U\over\pi}\int_0^{\infty}dyF_{\infty}^2(
iy){y^2\over\Omega^2+y^2}]\cr
&-&{U\over\pi}\int_0^{\infty}dyF_{\infty}^2(iy){\Omega^4\over(\Omega^2+y^2)^2}
\}
\label{eq: gammadef}
\end{eqnarray}
in the square-well approximation \cite{reviews} with $\omega_c$ the
cutoff frequency.  The transition temperature is determined in the standard
fashion
\cite{mcmillan,reviews,wu}
\begin{eqnarray}
&T&_c/t^*=\exp[-{1\over|U|\rho(0)}] \times\exp
[-{2\over\sqrt{\pi}}\int_0^{\infty}dyF_{\infty}^2(iy)]\times
\exp [-2\int_0^{\infty}dy{\rho(y)\over\rho(0)}{\Omega\over(\Omega+y)^2}]\cr
&\times&\{0.85\exp [-{2\over\pi}\int_0^{\infty}{dy\over
y}{\rho(y)\over\rho(0)}\tan^{-1}{y\over\omega_c}]\}\times
\exp
[{2\over\sqrt{\pi}}\int_0^{\infty}dyF_{\infty}^2(iy){\Omega^2\over\Omega
^2+y^2}\{1+{1\over 2}{\Omega^2\over \Omega^2+y^2}\}]
\label{eq: tceqn}
\end{eqnarray}
including all nonvanishing factors in the limit $|U|\rightarrow 0$.  The first
factor is the contribution to lowest order in $|U|$.  The remaining terms are
constant factors arising from the following four sources (in order of
appearance):  the phonon self-energy contributions; the electronic
self-energy contributions;
the finite bandwidth contributions; and the vertex corrections.
The phonon self-energy term (equal to $\exp[2.493]$ in
infinite dimensions)
is normally incorporated into the dressed phonon
propagator in ME theory $\lambda\equiv |U|\rho(0)
/[1-2.493|U|\rho(0)]$.  The electron self-energy
term approaches $e^{-1}$ as $\Omega\rightarrow 0$ and $e^0$ as
$\Omega\rightarrow \infty$.  The finite-bandwidth term should be
proportional to $\Omega$ if the phonon frequency is much smaller than the
electronic
energy scale, but should be replaced by the electronic energy scale when
the phonon frequency becomes the largest energy scale.  For concreteness,
the cutoff
frequency is chosen to be $\omega_c=0.6\Omega$, so that this factor
approaches the correct limits of $0.69\Omega$ as $\Omega\rightarrow 0$
\cite{wu} and $0.85$ as $\Omega\rightarrow\infty$ \cite{vandongen}.  Finally,
the fifth term contains the vertex corrections which vanish as
$\Omega\rightarrow 0$ \cite{migdal} and approach $\exp[-3.739]$ as
$\Omega\rightarrow\infty$ \cite{vandongen}.  The above equation for $T_c$
has the correct limits as $\Omega\rightarrow 0$
$[T_c=0.69\Omega\exp\{-(1+\lambda)/\lambda \}]$ and as
$\Omega\rightarrow\infty$
$[T_c=0.24\exp\{-1/|U|\rho(0)\}]$.
The coefficient of $\exp[-1/|U|\rho(0)]$ in Eq.~(\ref{eq: tceqn}) is plotted in
Fig.\ \ref{fig:2}.  Note that the maximum occurs at $\Omega\approx
0.8t^*$ (approximately $20\%$ of the effective bandwidth), but the transition
temperature is weakly dependent upon $\Omega$ for larger values.
The ME form for the transition temperature, $T_c=1.14\omega_c\exp[-(1+\lambda)
/\lambda]$, can be fit to the above form in Eq.~(\ref{eq: tceqn}) by
renormalizing the electron-phonon interaction $\lambda$ and the cutoff
frequency $\omega_c$ (since the contributions from second-order terms
[phonon self energy and vertex corrections] simply modify the prefactor of the
$T_c$ formula), explaining its robustness for real materials calculations.

In order to make comparisons with the quantum Monte Carlo calculations
\cite{freericks}, the self-consistent perturbation theory is solved
numerically. An energy cutoff employing 256 positive Matsubara frequencies is
used, and the self-consistent equations are iterated until the largest
deviation of each $G(i\omega_n)$ is
less than one part in $10^8$.  At half-filling the instability always
lies in the
CDW channel.  The effective electronic bandwidth in infinite dimensions is
of order $4t^*$ and the phonon frequency is chosen to be approximately
one-eighth
of this electronic bandwidth $(\Omega=0.5t^*)$ as in the quantum Monte
Carlo simulations.

Both the first-order
and second-order conserving approximations are compared to the quantum
Monte Carlo data in Fig.\ \ref{fig:3}.  The transition temperature determined
in the
second-order approximation is found to be renormalized (by more than a
factor of two at
low temperatures) from the transition temperature calculated in the
first-order approximation.
The second-order conserving approximation deviates
\cite{cdw} from the Monte Carlo results
when $g$ is greater than $0.4t^*$.
Furthermore, the turnover of the transition temperature curve
as $g$ increases
is not reproduced by these finite-order conserving approximations.
Probably an infinite-order summation scheme (such as the fluctuation-exchange
approximation \cite{baym}) will be required to properly account for this
turnover.

In the high-frequency limit $(\Omega\rightarrow\infty)$, the Holstein model
maps onto an attractive Hubbard model.  In this limit many ``direct'' diagrams
cancel against ``exchange'' diagrams (there is no interaction between electrons
with the same spin) and the perturbation theory is easier to control.
Quantum Monte Carlo simulations have already been performed \cite{jarrell},
and weak-coupling expansions through second order (for transition
temperatures) have also been investigated
with different techniques \cite{vandongen,japanese}.

The relevant diagrams for the free energy of the Hubbard model through third
order are depicted in Fig.\ \ref{fig:4}.  The transition temperature in
the CDW channel
is plotted in Fig.\ \ref{fig:5} for the first through third-order
approximations and
for the quantum Monte Carlo simulations.  Note that the agreement with
the Monte Carlo data is excellent and that the weak-coupling expansions
appear to have an alternating character for the Hubbard model with odd orders
overestimating $T_c$ and even orders underestimating $T_c$.

Weak-coupling conserving approximations (that are valid for all values of the
phonon frequency) have been carried out for the Holstein (and Hubbard) model at
half-filling.  Calculations must be performed to second order in the
effective electron-electron interaction in order to produce the correct
limiting behavior as $U\rightarrow 0$.  Comparison was made with quantum Monte
Carlo simulations and excellent agreement was found for weak coupling.

Migdal's theorem states that, in the small-frequency limit, transition
temperatures may be determined with a first-order approximation (using
a dressed phonon propagator) because the vertex corrections vary as
$\Omega /t^*$.  This
result is easily seen in our framework, but the corrections to $T_c$
due to a nonvanishing phonon frequency can be large (when expressed in terms of
the
bare parameters).  Migdal-Eliashberg theory is, however, quite robust,
and the effects of a finite band-structure and vertex corrections
(to lowest order) can be incorporated by renormalizing the phonon frequency
and the interaction strength (in the superconducting channel).

Extensions of these results off of half-filling, comparison with other
techniques such as the fluctuation-exchange approximation \cite{baym} or the
iterated perturbation theory \cite{georgesandkotliar}, and an
investigation of Coulomb effects will be presented elsewhere.

We would like to thank E. Nicol
for many useful discussions.
This research was supported in part by the NSF under Grants No.
PHY89-04035 and DMR90-02492.

%\end{document}

\begin{figure}
  \caption{
First and second order contributions to the free energy in a conserving
approximation for the Holstein model.  The solid lines are dressed
electronic Green's functions and the wiggly lines are phonon propagators.
The spin indices $\sigma$ and $\tau$ are summed over.  }
  \label{fig:1}
\end{figure}

\begin{figure}
  \caption{
Constant factor in the analytic formula for the superconducting transition
temperature in Eq. (7) plotted as a function of phonon frequency.  Note
the constant term has a maximum at $\Omega\approx 0.8t^*$.  }
  \label{fig:2}
\end{figure}

\begin{figure}
  \caption{
Charge-density-wave transition temperature for the Holstein model
at half-filling with $\Omega=0.5t^*$ plotted against the
electron-phonon interaction $g$.  The first order (solid line), second order
(dashed line), and quantum Monte Carlo results
[15] (solid dots) are included.
  }
  \label{fig:3}
\end{figure}

\begin{figure}
  \caption{
First through third order contributions to the free energy in a conserving
approximation for the Hubbard model (the dashed line is the electron-electron
interaction $U$).  The spin index $\sigma$ is summed over.
  }
  \label{fig:4}
\end{figure}

\begin{figure}
  \caption{
Charge-density-wave (and superconducting) transition temperature for the
Hubbard model at half-filling plotted against the interaction strength $|U|$.
The first order (solid line), second order (dashed line), third order
(dotted line), and quantum Monte Carlo results [14]
(solid dots) are included.
  }
  \label{fig:5}
\end{figure}


\begin{thebibliography}{10}
\bibitem{cdw} It is more difficult to fit the CDW transition temperatures
into a ME scheme because the interaction has a much stronger
temperature and frequency dependence in the CDW channel.
\end{thebibliography}
\end{document}